# CASA: A Framework for SLO and Carbon-Aware Autoscaling and Scheduling in Serverless Cloud Computing

Sirui Qi, Hayden Moore
*Colorado State University*
Fort Collins, USA
{alex.qi, hayden.moore}@colostate.edu

Ninad Hogade, Dejan Milojicic, Cullen Bash
*Hewlett Packard Labs*
Milpitas, USA
{ninad.hogade, dejan.milojicic, cullen.bash}@hpe.com

Sudeep Pasricha
*Colorado State University*
Fort Collins, USA
sudeep@colostate.edu

*Abstract—* Serverless computing is an emerging cloud computing paradigm that can reduce costs for cloud providers and their customers. However, serverless cloud platforms have stringent performance requirements (due to the need to execute short duration functions in a timely manner) and a growing carbon footprint. Traditional carbon-reducing techniques such as shutting down idle containers can reduce performance by increasing cold-start latencies of containers required in the future. This can cause higher violation rates of service level objectives (SLOs). Conversely, traditional latency-reduction approaches of prewarming containers or keeping them alive when not in use can improve performance but increase the associated carbon footprint of the serverless cluster platform. To strike a balance between sustainability and performance, in this paper, we propose a novel carbon- and SLO-aware framework called *CASA* to schedule and autoscale containers in a serverless cloud computing cluster. Experimental results indicate that *CASA* reduces the operational carbon footprint of a FaaS cluster by up to 2.6× while also reducing the SLO violation rate by up to 1.4× compared to the state-of-the-art.

*Keywords—serverless computing, container autoscaling, container scheduling, carbon emissions, service level objectives*

## I. INTRODUCTION

To prevent the temperature of the earth from exceeding the tipping point of runaway global warming, governments across the globe signed the Paris Agreement in 2016 and carbon tariffs have since been in place across several regions. However, the increasing carbon footprints of datacenters around the world are jeopardizing such efforts. Datacenters worldwide have been shown to consume more than 1% of global electricity production, which equals the electricity usage of an entire mid-size developed country [1]. This consumption is expected to grow to 3-13% of global electricity demand by 2030 [2]. Datacenters are also responsible for 2-4% of all global carbon emissions today and these continue to grow, with the increasing investments in AI and bitcoin mining [3]. Thus, reducing the carbon emissions of datacenters has taken on great urgency.

In recent years there has been a shift in cloud computing with the increased adoption of the serverless computing paradigm and approximately 50% of global enterprises have embraced this approach [4]. In serverless computing, cloud infrastructures (e.g., server nodes, virtual machines) are hidden from users and managed by the cloud service providers. Serverless computing has thus enabled the function-as-a-service (FaaS) paradigm, where developers focus on implementing fine-grained pieces of code (functions) that are packaged independently and transparently (to the developer) in lightweight virtualized containers and hosted in the cloud. FaaS provides an event-driven platform where functions are triggered by a specific event such as HTTP requests in user applications. In contrast to traditional cloud computing where developers reserve server nodes and inflexibly pay for bulky virtual machines (VMs), FaaS allows them to pay less through a pay-as-you-go model. Moreover, service providers can reduce carbon emissions by turning off idle containers [5]. This is significant because about 50% of energy in today's cloud datacenters is consumed by idle resources [1].

Unfortunately, FaaS does not necessarily reduce carbon emissions when compared to traditional cloud computing. This is because FaaS involves frequent container activities, such as initializing, starting up, shutting down, and scaling up or down. All aspects of such activities have notable carbon emission overheads, which get added to the total contribution from the execution of serverless functions. If cloud service providers alter the behaviors of containers in FaaS clusters to meet sustainability goals, e.g., by shutting down idle containers, it can have the negative effect of reducing performance, e.g., due to needing to cold-start the previously shut down containers, which requires the container image and associated library to be moved from persistent storage nodes to computing nodes, thus increasing end-to-end latency of requests.

Today's commercial serverless platforms such as AWS Lambda [6] Azure Functions [7], OpenWhisk [8], and OpenFaaS [9] allow hosting short-running functions with developer-specified service level objectives (SLOs). The SLOs typically require a service provider to successfully execute associated functions before a preset (latency) deadline such that service quality is maintained for users of a cloud service. These SLOs are customarily expressed as an *SLO violation rate* constraint, calculated as the proportion of serverless function executions (e.g., 5%) that are allowed to violate deadlines per request type. To satisfy SLO constraints, many prior works have proposed to "prewarm" containers (i.e., launch containers in advance) before external requests arrive and keep idle containers running in a cluster, in anticipation of new requests being able to use them with low latency [10] [11]. Such solutions can reduce the end-to-end latency and SLO violation rate. However, such prewarming and keeping alive techniques increase a cluster's utilization which contributes to increased operational carbon emissions.

In summary, optimizing the carbon footprint in serverless computing requires container scheduling and autoscaling frameworks to consolidate and reduce containers so that underlying resources are utilized in a carbon-efficient manner. At the same time, SLO optimization requires the frameworks to spread out and increase container resource allocations so that cold-start latencies can be reduced, and contention-related delays can be avoided, but this increases operational carbon overheads. To co-optimize these two conflicting objectives (carbon emissions and SLO violation rate), we propose a novel dual-objective framework for scheduling and autoscaling in FaaS clusters. The novel contributions of our work are:

- We propose a novel scheduling and autoscaling framework called *CASA* that combines local search and heuristic-switching techniques to co-optimize the dual objectives of carbon emissions and SLO violation rates;
- We comprehensively model the carbon emissions, container overheads, and SLOs in a FaaS computing platform;
- We compare *CASA* with three state-of-the-art serverless scheduling and autoscaling frameworks to show that *CASA* outperforms them in terms of carbon emissions and SLO violation rate, at different problem scales.

## II. RELATED WORK

Serverless computing is an appealing paradigm for both cloud providers (who can streamline their services more flexibly) and developers (who can better manage costs with the pay-as-you-go model). Several prior efforts have focused on improving the performance (latency, throughput, or SLO ) of serverless computing in datacenters [10] [12] [13] [14] [15] [16]. These works typically propose techniques that can efficiently perform container scheduling (i.e., mapping containers to servers spatially and temporally) and/or autoscaling (i.e., increasing and decreasing active containers to handle incoming function execution requests). The autoscaling of containers can be performed either horizontally or vertically, which involves adding containers to different nodes in the cluster, or adding containers to the same node, respectively.

Ebrahimpour et al. [10] proposed a heuristic-based scheduler that reduced the container cold-start latency (i.e., the latency to initialize a new container in response to requests). Using a dynamic waiting-time adjustment technique, their approach reduced the computing and memory resources spent on starting up new containers. A different heuristic approach proposed by Przybylski et al. [15] utilized a scheduler that prioritized less commonly invoked functions to reduce cold-start latency, such that cloud throughput was increased when compared to either a round-robin or first-in-first-out approach. Sinha et al. [12] proposed using an online machine-learning approach that can predict the resources needed to meet a function's SLO, after which it dynamically autoscales the containers for serverless requests. They showed an improvement in meeting each function's SLO over static-allocation approaches. Another machine-learning approach was proposed by Li et al. [13], which utilized a Mondrian Forest (a variant of a random forest built on split Mondrian trees, which are a variant of a decision tree). They demonstrated an improvement in the end-to-end latency when compared to a baseline approach that adopted the most recently used container routing policy. Schuler et al. [14] proposed a reinforcement learning (RL) based autoscaler that adjusted the number of VMs available for scaling in the system, depending on the system throughput (requests per second). A different deep RL approach utilizing a scoring system was proposed by Yu et al. [16]. They demonstrated that by utilizing a deep RL approach, the function completion time can be reduced when compared to greedy schedulers and hashing-based schedulers. *All of these performance-first serverless schedulers lack the focus on energy efficiency, making it difficult to maintain sustainability goals.*

The current focus of sustainability optimization within serverless cloud datacenters is to optimize their energy usage [17] [18]. The framework in Rastegar et al. [17] utilized an online scheduler based on linear programming that increased the energy efficiency of the datacenter. This was able to outperform a fixed schedule and an adaptive average scheduling approach. Given a cluster of battery-operated and renewable-energy-powered nodes, Aslanpour et al. [18] proposed an energy-aware scheduler utilizing a real-time energy efficiency model to determine the best node location for container placements. By doing so, they reduced the energy consumption of a cluster while preserving SLOs when compared to the default Kubernetes scheduler. However, *energy consumption does not always correlate with the carbon emissions of datacenters because other carbon-related factors must be considered such as the carbon intensity of the energy grid* [19]. By directly optimizing the carbon emissions of serverless datacenters, we believe that better sustainable serverless schedulers can be designed.

Carbon-aware scheduling within datacenters has been studied within traditional serverfull cloud computing [20] [21]. However, few efforts have so far explored carbon-aware serverless scheduling. Chadha et al. [1] proposed a carbon-aware scheduler for allocating functions across geo-distributed datacenters. They demonstrated a reduction in carbon emissions compared to the default Kubernetes scheduler. However, the latency in migrating short-lived functions between datacenters can be prohibitively high and is a limitation of relying on the geo-distribution strategy to reduce carbon emissions.

Our work addresses the challenge of realizing a sustainable serverless computing environment within a FaaS cluster via a novel dual-objective optimization framework that not only directly optimizes carbon footprint but also ensures that SLO constraints are respected as part of the scheduling and autoscaling strategy.

## III. SYSTEM MODEL

Our proposed *CASA* framework not only decides the number and location of containers for each function type execution request independently but also autoscales the CPU and DRAM resource allocations of each container in the cluster. *CASA* supports multiple containers being co-located on the same server node during autoscaling. We further characterize energy costs, operational carbon emissions, SLO violation rates, and average system load (i.e., utilization) of the FaaS cluster. Fig. 1 illustrates a FaaS cluster's infrastructure modeled in this work. The cluster is comprehensively modeled in terms of its power usage, carbon emissions, intra-cluster latency, and workload, as discussed in the rest of this section.

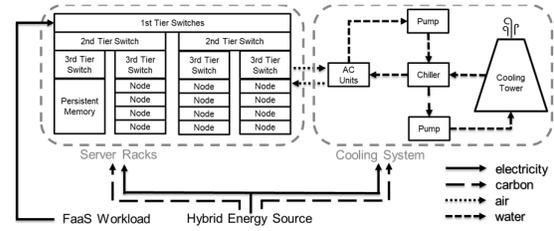

Fig. 1. Electricity/Carbon/Air/Water flow in a FaaS cluster

### A. Cluster Model

We consider a FaaS cluster consisting of $N$ computing nodes that are connected through a three-tier intra-cluster network, which introduces heterogeneous bandwidths from the persistent memory (storage nodes) to the computing nodes. Air conditioning (AC) units are used to cool the cluster and their overheads are also considered.

#### 1) Power Consumption

We divide the power consumption of a FaaS cluster into the total information technology (IT) power $P_{IT}$ and cooling power $P_{Cooling}$ [22]. $P_{IT}$ is the cumulative power $P_i$ of $N$ computing nodes, as well as power for storage nodes $P_S$ and networking components $P_N$.

$$P_{IT} = \sum_{i=0}^{N} P_i + P_S + P_N \quad (1)$$

We assume a fixed storage and networking power overhead. To understand the computing node-level power usage $P_i$ due to co-located containers, we performed CPU stress tests with different resource usage on a 128-core AMD computing node (EPYC 7713) and created a polynomial regression model, shown in Fig. 2. The relation between the $P_i$ and CPU core usage $U_i$ was approximated as:

$$P_i = AU_i^4 + BU_i^3 + CU_i^2 + DU_i + E \quad (2)$$

where $A, B, C, D, E$ are node-specific parameters. We observed a similar polynomial regression model across node types with different parameter values. For our work, we assume only one node type in the cluster and the model in (2) applies to all nodes. The CPU usage $U_i$ consists of the core usages from all co-located containers.

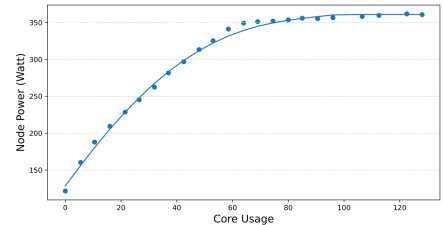

Fig. 2. Power consumption of EPYC 7713 for different numbers of cores used. Dots are real measurements, and the line shows the regression model.

The efficiency of the cooling system in Fig. 1 is under the influence of the outdoor temperature [20] [23], and thus the relation $P_{Cooling}$ is calculated as:

$$P_{Cooling} = E_t \times P_{IT}, t \in [0, \cdots, 23] \quad (3)$$

where $E_t$ represents cooling efficiency at an hour $t$ of solar time during the day. A smaller $E_t$ value indicates less $P_{Cooling}$ is used to remove the same amount of heat generated by $P_{IT}$.

#### 2) Carbon Emission and Energy Cost

To create a realistic carbon model and energy cost model, we assume that the cluster is powered by grid electricity that depends on mixed energy sources, e.g., from combustion generators, nuclear reactors, hydro generators, solar panels, and wind turbines. Due to the proportions of all energy sources varying hourly at a location, the corresponding carbon intensity and energy cost changes as well. We study the related data of Tampa, FL, USA in the summertime and plot the carbon intensity $CI$ together with electricity pricing $EP$ at different hours $h$ in Fig. 3. The cumulative carbon emissions $CA$ and energy cost $CO$ over a day can be calculated as:

$$CA = \sum_{h=0}^{23} CI_h \times (P_{IT,h} + P_{Cooling,h}) \quad (4)$$

$$CO = \sum_{h=0}^{23} EP_h \times (P_{IT,h} + P_{Cooling,h}) + CO_{water} \quad (5)$$

where $CO_{water}$ represents carbon emissions attributed to potable water production and wastewater treatment, the value of which depends on the quality of local water sources and local water treatment procedures. We use the water factor $I_t$ from [20] at an hour $t$ to calculate the amount of water-related carbon emissions at $t$ due to the cooling efforts $P_{cooling}$ as shown in:

$$CO_{water} = I_t \times P_{Cooling}, t \in [0, \cdots, 23] \qquad (6)$$

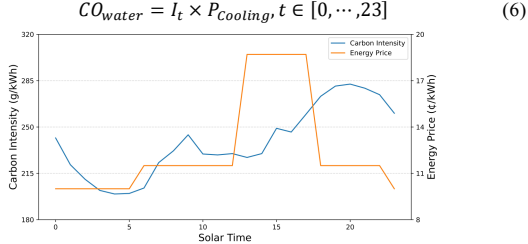

Fig. 3. Carbon intensity and energy price at Tampa, FL over solar time [24]

*B. SLO Model*

To quantify the SLOs in FaaS platforms, we first investigate the arrival pattern and request intensity of three real-world serverless production traces (Alibaba [25], Huawei [26], and Azure [27]). We find that all three serverless production traces have similarities in the request intensity and arrival patterns. An analysis based on the Azure trace [27] is discussed below. We use this trace to illustrate how each container works and impacts the SLO. By combining knowledge from the production traces and container operations, we introduce deadline laxity and the SLO violation rate model to quantify SLOs.

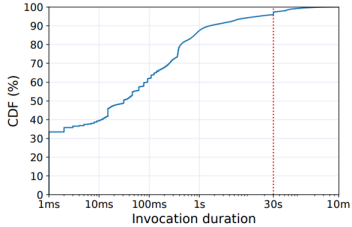

Fig. 4. Cumulative distribution function of functions' duration in Azure [27]

*1) Workload*

A research team cooperated with Azure to monitor its serverless function usage for 2 weeks in 2019 [27]. They found that more than 90% of serverless functions last less than 30 seconds as shown in Fig. 4. Expanding on that, we further investigate the arrival patterns and request intensity of this trace in 15-minute epochs. From Fig. 5, we observe that: 1) Request intensity of neighboring epochs changes from 0× to 13×; and 2) The number of unique function types changes from 13 to 62 in a 15-minute epoch (424 unique function types in total). From the data, we can infer that the request intensity of a serverless trace changes rapidly over short epochs which can introduce up to 13× more requests within 15 minutes. Further, unlike in many serverfull scenarios, the arrival pattern of function execution requests is not stable; up to 62 new function types invoke 62 unique functionalities within an epoch as shown in Fig. 5. Similar rapidly changing patterns are also observed in the function arrival patterns in [25] and [26]. These observations motivate us to develop a scheduling and autoscaling framework that can make real-time decisions and cope with fast-changing function workloads.

The studies in [25], [26], and [27] all indicate that most FaaS requests invoke short-lived functions requiring minimal resource usage, such as small data processing. Hence, in our workload model, we assume that each request in FaaS occupies the same small amount of CPU and DRAM resources in our environment to finish the associated task within its average execution time. The average execution time per function type is calculated based on the corresponding execution times over the two-week Azure trace [27]. A function ID represents a unique execution functionality that is shared by multiple requests of the same ID, all of which can be processed by the same container in parallel. Each container only accepts parallel requests for a single ID because of the container's unique image file and libraries that are needed for executing the functionality of that specific function ID. To support parallel computing sustainably, we allow containers to reconfigure their resource limits on the fly. This feature has been feasible in the Kubernetes experimental release [28], and we allow containers to reconfigure (autoscale) their sizes during an epoch, e.g., scale up from an initial allocation of 2 cores and 150MB DRAM to 4 cores and 300MB DRAM, to handle more parallel requests with the same function ID. An epoch is a custom duration of time (e.g., 15 minutes) where container scheduling decisions are made at the start of an epoch, and autoscaling can occur anytime within an epoch.

Lastly, we observe that additional resource utilization and latencies are incurred when a container is starting up, being idle, or shutting down, through our experiments on the Knative platform. These extra resource utilization and latencies are empirically modeled across function IDs and discussed in Section $VI.A$.

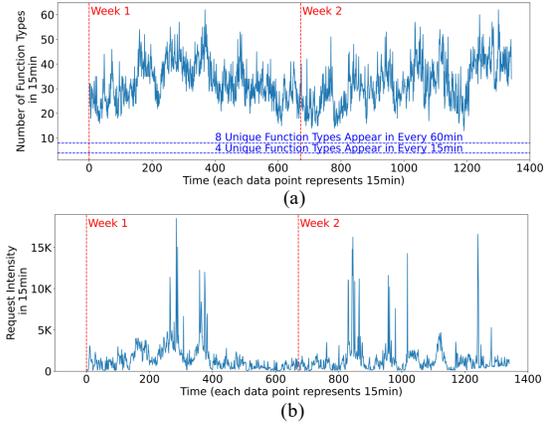

Fig. 5. (a) Request intensity and (b) arrival pattern of Azure trace [27]

*2) SLO*

A deadline is defined as the maximum time that an SLO allows for a request to be processed after its arrival at $T_{arrival}$. If the request does not finish execution after $T_{deadline}$, it will be dropped by the cluster and considered an SLO violation. After knowing the request's arrival time $T_{arrival}$ and deadline $T_{deadline}$, a deadline laxity can be calculated with function ID average execution time $T_{ave}$:

$$Laxity = (T_{deadline} - T_{arrival})/T_{ave} \qquad (6)$$

The larger the deadline laxity is, the more time a request is allowed to wait at the computing nodes or be migrated to a different node. Once the request stays in the cluster longer than $Laxity \times T_{ave}$, it is considered an SLO violation. Hence, the function ID-specific SLO violation rate $SL_{f,e}$ at epoch $e$ is defined as:

$$SL_{f,e} = V_{f,e}/N_{f,e} \qquad (7)$$

where $V_{f,e}$ represents the number of violations of function ID $f$ and $N_{f,e}$ represents the number of function ID $f$ requests, in epoch $e$.

*3) Cold-Start Latency*

As shown in Fig. 1, we assume that there is persistent memory that stores all the necessary container images and associated libraries. When there is a new function execution request, the corresponding container image for that function and required libraries will be copied from the persistent memory to the memory in the computing nodes, and the latency required for this is called the cold-start latency $T_{cold}$. The three-tier network in Fig. 1 introduces heterogeneous network bandwidths between the persistent memory and computing nodes. If the computing node is farther away from the persistent memory nodes, it means that data needs to go through more network switches which increases cold-start latency, Furthermore, there can be hundreds of container images and libraries transferred through the network concurrently. Hence, the network bandwidth to each node can be oversubscribed if too many container images are being sent to the same node. Considering the container image data footprint of co-located cold-start containers $C_{cold}$, network bandwidth of the node $B_{node}$, network switch delay $T_{switch}$, and the number of switch hops $NH$, cold-start latency $T_{cold}$ can be expressed as follows:

$$T_{cold} = C_{cold}/B_{node} + NH \times T_{switch} \qquad (8)$$

After modeling cold-start latency, the actual finish time of a request can be expressed as a function of the waiting time $T_{wait}$, cold-start latency $T_{cold}$, and average execution time $T_{ave}$:

$$T_{finish} = T_{arrival} + T_{wait} + T_{cold} + T_{ave} \quad (9)$$

where waiting time $T_{wait}$ is introduced when the assigned computing node is fully occupied, and the corresponding request must wait.

## IV. PROBLEM FORMULATION

We consider a FaaS provider that is managing incoming serverless function requests across multiple nodes. There are $Z$ number of unique function IDs that request different functionalities in the FaaS cluster. At the beginning of each epoch, the framework receives a forecast of the workload information including the function ID list $F = [f_i]_{i \in Z}$ and the corresponding request intensity list $R = [r_i]_{i \in Z}$ in the upcoming epoch. The requests can arrive at any time during the epoch. In each epoch, the framework is responsible for providing scheduling and autoscaling plans that include deciding: *(i)* the number of containers per function ID, *(ii)* locations of all containers on available compute nodes, *(iii)* request distributions to all containers, and *(iv)* node resource allocations for each container. The goal of the framework is to co-optimize two objectives: cumulative operational carbon emissions $CA_{cum}$ of the cluster and average SLO violation rate $SL_{ave}$ of $Z$ function IDs, while ensuring that the SLO violation rate constraint $Cstr$ is met.

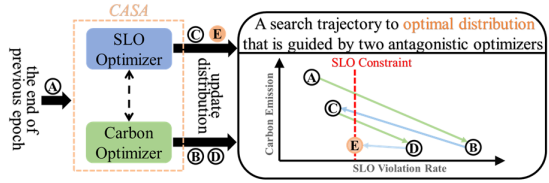

Fig. 6 *CASA* framework overview

## V. FRAMEWORK

Our *CASA* framework integrates an SLO optimizer that utilizes a self-guided local search with an SLO reduction objective and a carbon optimizer that utilizes a self-guided local search with a carbon reduction objective. As shown in Fig. 6, the two optimizers are antagonists in their optimization directions. The SLO optimizer determines the number of container instances and distributes them across the cluster to reduce cold-start latency and minimize resource contention. Therefore, decisions from the SLO optimizer will likely reduce the SLO violation rate, but the increased resource utilization increases carbon emissions. Meanwhile, the carbon optimizer decreases the number of containers and consolidates them in the cluster. Fewer containers reduce energy consumption but can increase SLO violations due to increased contention. In this section, we discuss how *CASA* utilizes both optimizers and switches between them to jump out of local optima to enable carbon-efficient container management while also reducing the SLO violation rate.

### A. Dual-Objective Optimization

Fig. 7 shows an overview of the algorithmic flow in the *CASA* framework. At the beginning of an epoch $e$ (① in Fig. 7), the iteration ceiling $gen$, a function ID list $F_e$, a request intensity list $R_e$, previous container distribution $D_{e-1}$, current carbon intensity $CI_e$, current energy pricing $EP_e$, the SLO violation rate constraint $Cstr$, the number of nodes $N$ in a cluster, iteration limit $K$ for local search in optimizers, and current epoch $e$ are input into *CASA*. $F_e$ contains all function IDs $f$ that appear in an epoch $e$ and $R_e$ consists of the corresponding request intensity of each $f$. The framework also inherits the container distribution $D_{e-1}$ from the end of the previous epoch $e-1$. Two empty blacklists $B_{SL}$ and $B_{CA}$ are initialized for optimizer usage. The initial carbon emissions $CA$, the average SLO violation rate $SL$, the energy cost $CO$, and the average load $LO$ of the cluster are calculated based on detailed simulation of the container allocations on the target serverless cluster.

The iteration ceiling $gen$ is preset to ensure that the framework outputs a container distribution plan within the desired time (②). This is because *CASA* is expected to make real-time decisions in short epochs without delaying incoming serverless requests.

*CASA* monitors if the current distribution plan $D_e$ will satisfy the preset SLO constraint $Cstr$ (③). If $D$ is not estimated to satisfy the preset constraint, *CASA* will utilize the SLO optimizer repeatedly to update $D_e$ to reduce the SLO violation rate $SL$. However, once $D_e$ can satisfy the constraint, *CASA* switches to the carbon optimizer to further reduce carbon emissions. Importantly, *CASA* can switch back to the SLO optimization if the new distribution $D_e$ from the carbon optimizer violates the SLO constraint. After $gen$ is crossed, *CASA* is expected to maintain a precise balance between the SLO violation rate and carbon emissions while providing a low-carbon container schedule that satisfies the SLO constraint.

Furthermore, the switching back and forth between the two optimizers helps *CASA* jump out of the local optima of both the SLO violation rate and carbon emissions, which forces *CASA* to explore unknown decision spaces for the global optima. Hence, the switch mechanism mitigates the problem where local search can become trapped in local optima. Details of the SLO optimizer and the carbon optimizer are discussed in the following subsections.

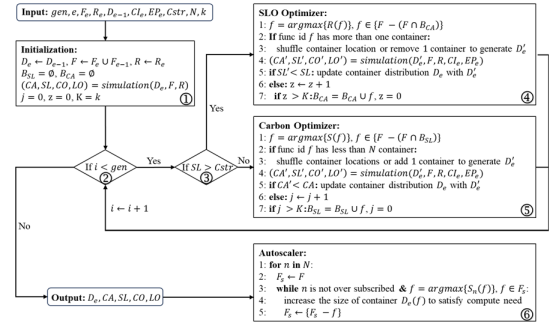

Fig. 7 *CASA* dual-objective optimization algorithmic flow

### B. SLO Optimization

To help the SLO optimizer reduce the SLO violations quickly, function IDs that contribute to more SLO violations $S(f)$ will be searched earlier (line 1 in ④). The local search order is not fixed as the SLO violations of unsearched IDs will be updated after each local search due to the co-location effect. During a local search, the SLO optimizer not only adjusts the container number of searched IDs but also tries to shuffle the current container distribution of searched IDs (lines 2-3 in ④). Shuffling can help containers get away from over-subscribed nodes and therefore acquire more CPU and DRAM resources for parallel computing. The SLO optimizer will update the current container distribution if its container adjustments can help reduce the average SLO violation rate instead of just a single-ID's SLO violation rate (lines 4-5 in ④).

To prevent the SLO optimizer from performing local searches on invalid search points, the SLO optimizer keeps a blacklist $B_{SL}$. The blacklist is empty at the beginning of each epoch but starts recording function IDs that are not reducing the SLO violation rate after $K$ rounds of local searches. If a function ID is on the blacklist, the SLO optimizer will avoid searching new distributions for this ID and switch to searching the other IDs (lines 6-7 in ④).

### C. Carbon Optimization

Different from the SLO optimizer, function IDs that have higher request intensity $R(f)$ will be searched earlier in this optimizer (line 1 in ⑤). This is because the carbon optimizer considers IDs with a higher request intensity to be more likely to consume more energy and therefore release more carbon emissions into the atmosphere. Such local search order is fixed until the blacklist $B_{CA}$ interrupts it.

During a local search, the carbon optimizer not only decreases the container number of searched IDs but also tries to shuffle the current container distribution of searched IDs (lines 2-3 in ⑤). Shuffling can help consolidate the containers in a cluster which helps reduce carbon emissions. The carbon optimizer will update the current container distribution as long as the new decisions $D'$ with current $CI_e$ can reduce overall carbon emissions $CA$ (lines 4-5 in ⑤) instead of energy consumption [17] [18]. The value of $CI_e$ decides the sensitivity of a cluster's $CA$ due to changes in $D$.

Similar to the SLO optimizer, the carbon optimizer keeps its own blacklist $B_{CA}$, to avoid repeated local searches on one function ID (lines 6-7 in ⑤). Such a blacklist design can help the optimizer jump out of local optima and explore more function IDs' distribution plans.

*D. Autoscaling*

After the container schedule (distribution) has been created and the epoch commences, our framework supports run-time autoscaling of containers in nodes (lines 1-2 in ⑥) to handle request spikes that may exceed an existing containers current CPU and DRAM allocation. The container for the function ID with the highest number of requests is prioritized for scaling to avoid resource contention and potential SLO violations. The autoscaling in node $n$ stops when the node is fully subscribed (i.e., has 100% utilization, lines 3-5 in ⑥).

## VI. EXPERIMENTS

*A. Experiment Setup*

We compared our *CASA* framework with three state-of-the-art approaches: an improved Kubernetes-based default scheduler using a scoring system (*Score*) [29], a hybrid algorithm that used both containers and virtual machines (*Hybrid*) [13], and a Q-learning based approach (*RL*) [14]. For CASA we empirically found the best value of K = 5. All frameworks provide real-time decisions at the start of each 15-minute epoch. We assume that the forecast of function arrival for the upcoming epoch arrives at least 3 minutes before the epoch starts and thus each framework has 3 minutes to generate its decisions. We developed and validated a custom Python simulator for FaaS clusters that integrates our models discussed earlier, and within which we implement all frameworks to work with our FaaS cluster environment. The function-based workload used to evaluate all frameworks is based on the Azure FaaS cloud trace [27].

We assumed that the baseline container resource utilization for each function ID will occupy 2 CPU cores and 150MB DRAM and is autoscaled up in multiples of this allocation. Shutting down containers takes 15 seconds based on our experiments on the open-source Knative platform. The compute node hardware we consider is an AMD EPYC 7713 server (same as that used in the Knative experiments) which has 128 cores and 64 GB of DRAM. For our first experiment, we use a baseline configuration with request intensity R=20× of the Azure trace, number of nodes N=4, and SLO violation rate constraint Cstr=5%. In later experiments, we explore the impact of altering the request intensity from 5× to 40×, the deadline laxity from 10× to 20×, Cstr from 5% to 10%, and CPU nodes N from 2 to 32. To comprehensively evaluate *CASA*'s performance, we use four metrics: cumulative energy cost of the cluster, average SLO violation rate of all function IDs, cumulative (operational) carbon emissions of the cluster, and average system load (utilization) of the cluster.

*B. Experiment Results*

*1) Comparison with State-of-the-Art in Baseline Configuration*

In Fig. 8, the cumulative results are shown for our baseline configuration. *CASA* can be seen to outperform the state-of-the-art frameworks in all four metrics. *CASA* and *Score* both reach the lowest SLO violation rate. But *CASA* outperforms *Score* in terms of energy cost, carbon emissions, and system load by 1.9×, 1.9×, and 1.6× respectively. These results indicate that it is feasible to co-optimize the performance (SLO) and sustainability (carbon) in FaaS clusters with a framework like *CASA* without losing performance to performance-focused frameworks such as *Score*.

*2) Sensitivity Analysis on Request Intensity*

From Section *III*, we observed that the request intensity and arrival pattern change rapidly in a serverless workload. In this experiment, we explored how all frameworks behave across different request intensities (from 5× to 40×), with all other aspects remaining unchanged from the baseline configuration. Fig. 9 shows the results of this experiment. The 5× request intensity in Fig. 9. represents an under-subscribed situation, where *CASA* can maintain a preset 5% SLO violation rate while focusing on carbon emission reduction. For this intensity, *CASA* outperforms *Score*, *Hybrid*, and *RL* by 2.2×, 2.8×, and 1.9× respectively. Notably, *Score* provides a 3% lower SLO violation rate than *CASA* in the under-subscribed 5× request intensity case. This is because *CASA* has a preset 5% SLO violation rate goal and puts more effort into carbon reduction once the constraint is satisfied. Under higher request intensities such as 10×, 20×, and 40×, *CASA* not only provides the lowest SLO violation rate and carbon emissions but also outperforms the best state-of-the-art *Score* by 3× in SLO violation rate and 1.36× in carbon emissions. This is because *CASA* prioritizes SLO optimization until satisfying the preset SLO constraint under high request intensity scenarios.

*CASA*'s reductions in carbon translate to lower costs and system loads. Compared to *Hybrid*, *Score*, and *RL*, *CASA* reduces the system load by up to 3.9×, 2.1×, and 1.6× respectively and energy costs by up to 2.8×, 2.1×, and 1.9× respectively. A lower system load can help clusters be more resilient to request peaks and lower energy costs enable affordable operation budgets for service providers.

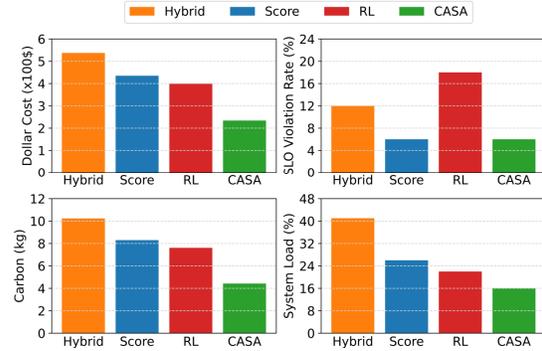

Fig. 8. Comparing the cost, SLO violation rate, carbon emission, and system load of all FaaS frameworks for baseline configuration.

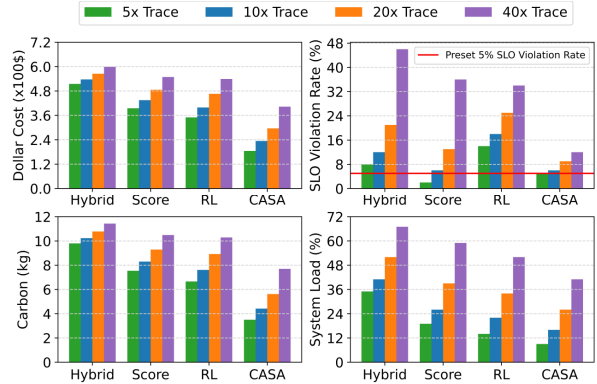

Fig. 9. Comparing the cost, SLO violation rate, carbon emission, and system load of all FaaS frameworks on different request intensities.

*3) Sensitivity Analysis on Deadline Laxity and SLO Constraint*

To evaluate the frameworks' ability to maintain the same SLO violation rate no matter how tight or relaxed the deadline laxity is, we vary the deadline laxity of arriving function requests from 10× to 20× of the corresponding execution time and set a 10% SLO constraint, with all other aspects remaining unchanged from the baseline configuration. Fig. 10 shows the results of this experiment. *CASA* can be observed to be more flexible than other frameworks when handling a variety of deadline laxity goals, as can be observed in Fig. 10(a). *CASA*'s dual-objective optimization approach maintains a consistent 10% SLO violation rate over different deadline laxities, with an improvement of 2.4× over the *Score*, 3.8× over the *Hybrid*, and 3.8× over the *RL* framework. *CASA* is better able to maintain the performance goal (SLO) because of its frequent checks of the SLO violation rate constraint during its optimization process.

We further evaluate *CASA*'s ability to operate under different SLO violation rate constraints, by varying the preset SLO goal from 5% to 10%, with all other aspects remaining unchanged from the baseline configuration, and comparing it with the other frameworks. From Fig. 10(b), we can see that *CASA* can provide different solutions according to the configurable SLO goal. When the SLO goal is 5%, *CASA* puts more efforts on SLO optimization and sacrifices more carbon emissions than its solution with a 10% SLO goal. *CASA*'s ability to outperform other frameworks in both the SLO constraint and carbon emissions demonstrates the effectiveness of the dual-optimizer approach in *CASA* which can adapt to different SLO goals and still provide excellent solutions.

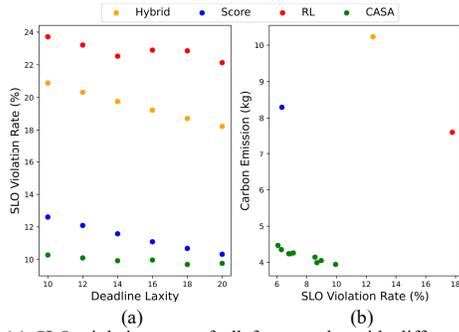

Fig. 10. (a) SLO violation rate of all frameworks with different deadline laxities and (b) carbon emissions of all frameworks with different SLO goals.

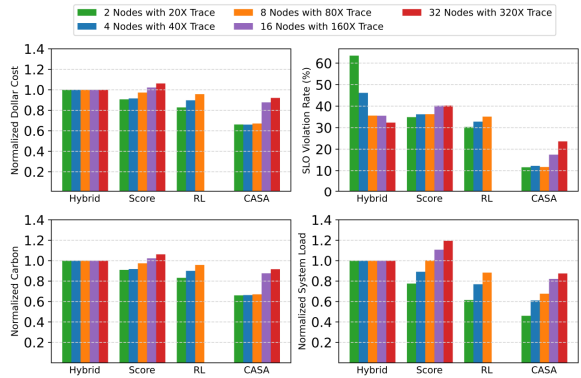

Fig. 11 Comparing the cost, SLO violation rate, carbon emissions, and system load of all FaaS frameworks on 2-, 4-, 8-, 16-, and 32-node clusters.

*4) Scalability Analysis on Cluster Size*

Lastly, we explore the scalability of our approach by increasing the cluster size from 2 to 32 nodes and request intensity from 20× to 320× accordingly, with all other aspects remaining unchanged from the baseline configuration. All frameworks are required to output their best solutions within 3 minutes no matter the problem size. The results for all frameworks are shown in Fig. 11. All normalized energy costs, carbon emissions, and system load values take the Hybrid framework as the baselines. We adjust the request intensity for different cluster sizes to generate sufficient system loads. From Fig. 11, we can observe that *CASA* always provides the lowest carbon emissions and SLO violation rate across different cluster sizes, where it outperforms the best state-of-art *RL* approach by up to 2.6× in SLO and 1.4× in carbon emissions. Note that the best state-of-art *RL* is unable to provide results in the 16-node and 32-node clusters. This is because the *RL* algorithm considered in that work utilizes a Q-table to store its policy, and the Q-table is not scalable to large cluster training in a fine-grained scheduling problem. The results in Fig. 11 indicate that *CASA* can scale up to a large design space with local search components. *CASA* learns from its local search history to avoid invalid search points (Fig. 7, lines 6-8 in ④ and ⑤) and therefore speeds up the optimization process, which helps it arrive at superior solutions for larger problem scales.

## VII. CONCLUSION

In this paper, we studied the problem of real-time scheduling and autoscaling in a FaaS cluster to co-optimize carbon emissions and the SLO violation rate without geographical task migration. We presented a novel framework called *CASA* for dual-objective optimization of real-time local FaaS scheduling and autoscaling. Our framework is able to adapt to real-world serverless workloads that contain fast-changing function request intensities and variable function arrival patterns. *CASA* was shown to have a better reduction in carbon emissions and SLO violation rates than the state-of-the-art, and also showed better scalability to variations in function request intensity, deadline laxity, SLO constraints, and cluster sizes. In our experiments, *CASA* realized up to a 2.6× carbon emissions reduction and up to a 1.4× reduction in the SLO violation rate compared to the best state-of-the-art FaaS scheduling framework. Thes results show the promise of *CASA* to realize truly sustainable serverless computing.

ACKNOWLEDGMENT

This research was supported by HPE and grants from the National Science Foundation (CCF-2324514, CNS-2132385).

REFERENCES

[1] M. Chadha, T. Subramanian, E. Arima, M. Gerndt, M. Schulz, O. Abboud, "GreenCourier: Carbon-Aware Scheduling for Serverless Functions," *ACM WoSC*, 2023.
[2] W. Hanafy, et al., "CarbonScaler: Leveraging Cloud Workload Elasticity for Optimizing Carbon-Efficiency," *ACM MACS,* 2023.
[3] Z. Cao, et al., "Toward a Systematic Survey for Carbon Neutral Data Centers," *IEEE CST*, 2022.
[4] M. Pandey, et al., "FuncMem: Reducing Cold Start Latency in Serverless Computing Through Memory Prediction and Adaptive Task Execution," *ACM SAC*, 2024.
[5] Y. Li, et al., "Serverless Computing: State-of-the-Art, Challenges and Opportunities," *IEEE TSC*, 2023.
[6] AWS Lamda, "https://aws.amazon.com/lambda/".
[7] Azure Functions, "https://azure.microsoft.com/en-us/".
[8] OpenWhisk, "https://openwhisk.apache.org/".
[9] OpenFaas, "https://www.openfaas.com/".
[10] H. Ebrahimpour, at al., "A heuristic-based package-aware function scheduling approach for creating a trade-off between cold start time and cost in FaaS computing environments," *ACM TJS*, 2023.
[11] R. B. Roy, et al., "Icebreaker: Warming serverless functions better with heterogeneity," *ACM ASPLOS*, 2022.
[12] P. Sinha, K. Kaffes, N. Yadwadkar, "Shabari: Delayed Decision-Making for Faster and Efficient Serverless Functions," *arxiv*, 2024.
[13] S. Li, at al., "GOLGI: Performance-Aware, Resource-Efficient Function Scheduling for Serverless Computing," *ACM SOCC*, 2023.
[14] L, Schuler, et al, "AI-based Resource Allocation: Reinforcement Learning for Adaptive Auto-scaling in Serverless Environments," *IEEE/ACM CCGrid*, 2021.
[15] B. Przybylski, et al., "Data-driven scheduling in serverless computing to reduce response time," *IEEE/ACM CCGrid*, 2021.
[16] H. Yu, et al, "FaaSRank: Learning to Schedule Functions in Serverless Platforms," *IEEE ACSOS*, 2022.
[17] S. Rastegar, et al., "EneX: An Energy-Aware Execution Scheduler for Serverless Computing," *IEEE TII*, 2024.
[18] M. Aslanpour, et al., "Energy-Aware Resource Scheduling for Serverless Edge Computing," *IEEE/ACM CCGrid*, 2022.
[19] M. A. B. Siddik, et al., "The environmental footprint of data centers in the United States," *Environmental Research Letters,* 2021.
[20] S Qi, D. Milojicic, C. Bash, and S. Pasricha, "SHIELD: Sustainable hybrid evolutionary learning framework for carbon, wastewater, and energy-aware data center management," *ACM IGSC*, 2023.
[21] S Qi, D. Milojicic, C. Bash, and S. Pasricha, "MOSAIC: A Multi-Objective Optimization Framework for Sustainable Datacenter Management," *ACM HiPC*, 2023.
[22] K. M. U.Ahmed, et al., "A review of data centers energy consumption and reliability modeling," *IEEE Access*, 2021.
[23] Q. Zhang, et al., "A survey on data center cooling systems: Technology, power consumption modeling and control strategy optimization," *Journal of Systems Architecture*, 2021.
[24] TECO, "tampaelectric.com/residential/saveenergy/energyplanner".
[25] A. Wang, et al, "FaaSNet: Scalable and fast provisioning of custom serverless container runtimes at alibaba cloud function compute," in *USENIX ATC*, 2021.
[26] A. Joosen,et al., "How does it function? characterizing long-term trends in production serverless workloads," *ACM SoCC*, 2023.
[27] M. Shahrad, et al., "Serverless in the Wild: Characterizing and Optimizing the Serverless Workload at a Large Cloud Provider," *USENIX ATC*, 2020.
[28] Kubernetes, "https://kubernetes.io/".
[29] Azure Traces, "github.com/Azure/AzurePublicDataset/tree/master".
[30] K. M. U.Ahmed, et al., "A review of data centers energy consumption and reliability modeling," *IEEE Access*, 2021.